\documentclass[%
reprint,
showpacs,
bibnotes,
amsmath,amssymb,
aps,
pra,
]{revtex4-1}

\usepackage[english]{babel}
\usepackage{graphicx}% Include figure files
\usepackage{dcolumn}% Align table columns on decimal point
\usepackage{bm}% bold math
\usepackage{hyperref}
\usepackage{multirow}
\usepackage{array}
\usepackage{booktabs}
\usepackage{ctable}
\usepackage{upgreek}
\usepackage{epsfig}
\usepackage{mathrsfs}
\usepackage{amssymb}
\usepackage{amsbsy}
\usepackage{color}
\usepackage{cancel}
\usepackage{pifont}
\usepackage{marginnote}
\newcommand{\bcen}{\begin{center}}
\newcommand{\ecen}{\end{center}}
\newcommand{\btab}{\begin{tabular}}
\newcommand{\etab}{\end{tabular}}
\newcommand{\bdes}{\begin{description}}
\newcommand{\edes}{\end{description}}

\newcommand{\beq}{\begin{equation}}
\newcommand{\eeq}{\end{equation}}
\newcommand{\bea}{\begin{eqnarray}}
\newcommand{\eea}{\end{eqnarray}}
\newcommand{\non}{\nonumber}

\newcommand{\half}{\frac{1}{2}}
\newcommand{\bary}{\begin{array}}
\newcommand{\eary}{\end{array}}
\newcommand{\benum}{\begin{enumerate}}
\newcommand{\eenum}{\end{enumerate}}
\newcommand{\bitem}{\begin{itemize}}
\newcommand{\eitem}{\end{itemize}}

\newcommand{\bgam}{\mbox{\boldmath $ \gamma $}}

\newcommand{\blam}{{\boldsymbol{\lambda}}}

\newcommand{\be} { \mbox{\boldmath $e$}}

\newcommand{\bk} { \bm{k} }

\newcommand{\bq} { \bm{q} }

\newcommand{\bzero} { {\boldsymbol{0}}}

\newcommand{\dou}{\partial}

\newcommand{\D}[1]{\mbox{d}{#1}}

\newcommand{\eqn}[1] {Eq.~(\ref{#1})}

\newcommand{\Fig}[1]{Fig.~\ref{#1}}

\makeatletter

\newcommand{\Rmnum}[1]{\expandafter\@slowromancap\romannumeral #1@}
\makeatother

\newcommand{\myfigwidth}{0.85\columnwidth}

\newcommand{\as}{a_{s}}

\newcommand{\kf}{k_F}

\newcommand{\rhop}{\rho_b}
\newcommand{\rhobc}{\rho_c}
\newcommand{\rhoF}{\rho_F}

\newcommand{\mylabel}[1]{\label{#1}} 

\newcommand{\mycite}[1]{\cite{#1}}

\newcommand{\FNkappa}{$\kappa=\frac{{(2 \pi)}^{4/3}}{4-\sqrt{2}} \approx 4.484$}

\newcommand{\FNlamNGbcs}{At a given temperature,  we have verified the expected result that $\lambda_{ng}(\as,T)$ increases for a smaller negative scattering length.}

\newcommand{\FNCalcEffort}{It is worth mentioning that, unlike in free vacuum, an analytical expression for the imaginary part of $M(\omega^+,\bq)$ is not available even for spherical gauge field. On top of this, the multiple integrals in evaluating $\rhop(T,\mu)$ and $\rhobc(T,\mu)$ which are challenging for an efficient numerics even in free vacuum, are extremely involved in presence of gauge fields due to various factors including principal valued integrals whose locations also have to be obtained numerically, sharp features and divergences in the integrand, among many.}

\newcommand{\OurTitle}{Fluctuation Theory of Rashba Fermi Gases}

\newcommand{\eqnFINALACTION}{A8}

\usepackage{lineno}
\begin{document}
\relax

%\preprint{}

%\preprint{}

%\title{Essential role of beyond-Gaussian effects and \\ the enhancement of superfluid $T_c$ of Rashba Fermi Gases}

\title{\OurTitle}

\author{Jayantha P.~Vyasanakere$^{1}$}
%\email{jayavyasa@gmail.com}
\author{Vijay B.~Shenoy$^2$}
\email{shenoy@physics.iisc.ernet.in}
\affiliation{$^1$Department of Physics, University College of Science, Tumkur University, Tumkur 572 103, India}
\affiliation{$^2$Department of Physics, Indian Institute of Science, Bangalore 560 012, India}



\date{\today}

\begin{abstract}

Fermi gases with generalized Rashba spin orbit coupling induced
 by a synthetic gauge field
 have the potential of realizing many interesting states such as rashbon condensates and topological phases. Here we develop a fluctuation theory of such systems and demonstrate that beyond-Gaussian effects are {\it essential} to capture the physics of such systems. We obtain their phase diagram by constructing an approximate non-Gaussian theory. We conclusively establish that spin-orbit coupling can enhance the exponentially small transition temperature ($T_c$) of a weakly attracting superfluid to the order of Fermi temperature, paving a pathway towards high $T_c$ superfluids.
\end{abstract}


\maketitle

Construction and study of model Hamiltonians with quantum gases has opened up the possibility of not only addressing long standing questions\cite{Ketterle2008,Bloch2008,Esslinger2010,Bloch2012,Trabesinger2012,
Cirac2012} but also creating systems that are not conventional. The recent advances in synthetic gauge fields\mycite{Dalibard2011,Lin2011,Williams2012,Liu2009,Shanxi2012,MIT2012}  have provided new impetus, motivating studies of interacting bosons and fermions in their presence that are of interest to a wide array of physicists (see review Ref.~\onlinecite{Goldman2014}). 

A uniform non-Abelian gauge field results in a 
generalized Rashba spin orbit coupling (RSOC).  
Interacting fermions with RSOC have many
interesting and novel
features,\mycite{Vyasanakere2011TwoBody,Vyasanakere2011BCSBEC,Yu2011,Hu2011}
while additional Zeeman fields can help realize topological
states.\mycite{Gong2011,Iskin2011a,Han2012} Even in the
presence of weak attractive interactions, a crossover from a
BCS type superfluid to a rashbon-BEC can be achieved by
increasing the strength of RSOC.\cite{Vyasanakere2011TwoBody,Vyasanakere2011BCSBEC}
Rashbon-BEC is a condensate of rashbons -- a bosonic
bound state of a fermion pair in  presence of large RSOC -- whose mass (between 2-2.5
fermion mass) determines its transition temperature which is of
the order of the Fermi temperature. Indeed early
studies\cite{Yu2011,Vyasanakere2011Rashbon} suggest this
enhancement of the transition temperatures of weakly attractive
systems by means of RSOC. In addition to
this, the rashbon condensate/gas has several uncommon and
unique traits -- for
example, unlike the case of the boson-boson interaction in the
usual BEC,\cite{Hu2006} the rashbon-rashbon interaction is {\it
  independent of the interactions between the constituent
  fermions}.\cite{Vyasanakere2012Collective} The hunt for the
`best possible $T_c$' in these systems, along with the novel
physics just noted, motivates this study to take the vital, if challenging, step
-- construction of a finite temperature theory including
fluctuations beyond the mean-field.

In this report, we develop, for the first time, a fluctuation theory of the normal state of an interacting Fermi gas with RSOC. We show that the Gaussian theory, which provides an excellent qualitative description of the BCS-BEC crossover in systems without  RSOC\cite{Nozieres1985,Drechsler1992,SaDeMelo1993,Randeria1995,Zwerger2012} is woefully inadequate to  describe the normal state of systems with RSOC {\em even at a qualitative level.} We develop an approximate theory, including the crucial beyond-Gaussian effects, and use it to obtain the phase diagram of interacting Rashba Fermi gases. Novel results include a clear demonstration of the enhancement of the superfluid transition temperature ($T_c$) in weakly attracting system from an exponentially small value to that of the order of Fermi temperature ($T_F$). We also show that in the regime of weak interactions, the superfluid transition temperature is a non-monotonic function of RSOC.

\noindent
{\bf Formulation:} Choosing units where the fermion mass $m$ and the Planck's constant $\hbar$ are unity, the kinetic energy of a spin-orbit coupled fermion with momentum $\bk$ in three spatial dimensions is
%\beq\mylabel{eqn:KineticEnergy}
$\varepsilon_{\bk \, \alpha} = \frac{k^2}{2} - \alpha |\bk_\lambda| + \frac{\lambda_m^2}{2}.$ 
%\eeq
Here $\alpha(=\pm 1)$ is the helicity, $\bk_\lambda = \lambda_x k_x \be_x + \lambda_y k_y \be_y + \lambda_z k_z \be_z$, where
$\blam \equiv (\lambda_x,\lambda_y,\lambda_z) \equiv \lambda \hat{\blam}$ describes a `vector' in the gauge field configuration space, and $\lambda_m=\text{Max}(\lambda_x,\lambda_y,\lambda_z)$. Non-Abelian gauge fields(equivalently RSOC) 
%(at least two components of  $\blam$ are non-zero)  
with high symmetry, such as the spherical gauge field with $\lambda_x = \lambda_y = \lambda_z = \lambda/\sqrt{3}$, are of particular interest. 
%where all the components of $\blam$ are equal, i.e., 
We refer to the absence of gauge fields/RSOC $(\blam=\bzero)$ as ``free vacuum''.

A finite density $\rho_0$ of fermions determines a characteristic momentum scale $k_F$ defined by $\rho_0 = \kf^3/3 \pi^2$ and an associated energy/temperature scale $E_F = T_F = k_F^2 / 2$. The singlet interaction (bare strength $\upsilon$) between the fermions is characterized by a scattering length $\as$. Physics at temperature $T$ and chemical potential $\mu$ with volume $V$ is studied using functional integral methods. After introducing pairing fields $\eta(q), q\equiv(iq_\ell,\bq)$ ($iq_\ell$-- Bose-Matsubara frequency, $\bq$ -- wave vector) and integrating out the fermions, the action 
\beq\mylabel{eqn:SetaF}
\textstyle{{\cal S}[\eta,F] = - \ln\det[-G^{-1}] - \frac{1}{\upsilon} \sum_q \eta^\star(q) \, \eta(q).}
\eeq
is obtained, where $F$ is a source field and $G$ is the Greens function (functional of $\eta$ and $F$).\footnote{\eqn{eqn:SetaF} is derived as Eqn.~\eqnFINALACTION~of Supplemental Material which may be referred for details.} To study the normal state physics, we expand the exact action \eqn{eqn:SetaF} about the saddle point where $\eta(q) = 0$,
%\beq
\bea 
\mylabel{eqn:SExpansion}
{\cal S} & \approx - \ln\det[-G_0^{-1}] - \frac{1}{\upsilon} \sum_q \eta^\star(q) \, \eta(q)  +  \sum_q \gamma^*(q) L(q) \gamma(q) \non \\
& \!\!\!\!\!\!\!\!\!\!\! + \sum_{q_1,q_2,q_3,q_4} \gamma^*(q_1)\gamma^*(q_2)K(q_1,q_2;q_3,q_4)\gamma(q_3)\gamma(q_4)
\eea
%\eeq
up to quartic order ($\gamma = \eta + F$); $G_0$ is the non-interacting Greens function. The quantities $L$ and $K$ are derivatives of the action (\eqn{eqn:SetaF}) to appropriate order in $\eta$. They are constrained by conservation laws, e.g., the arguments of $K$ have to satisfy momentum conservation.

\noindent
{\bf Gaussian Fluctuation Theory:} Retaining only the first three terms in \eqn{eqn:SExpansion} produces the Gaussian fluctuation theory\cite{Nozieres1985,Drechsler1992,SaDeMelo1993}, quadratic in $\eta$. Upon integration of the $\eta$ fields, we obtain
\beq\mylabel{eqn:Sgauss}
\textstyle{
{\cal S}_{\text{g}}[F] = - \ln\det[-G_0^{-1}] \! +  \! \sum_q \ln{M(q)} -\! \sum_q F^*(q)  \chi(q)  F(q),
}
\eeq
where $M(q) =L(q) - \frac{1}{\upsilon} = L(q) + \frac{1}{V} \sum_{\bk}\frac{1}{{|\bk|}^2} - \frac{1}{4 \pi \as} $,
%\beq\mylabel{eqn:Lofq}
$L(q) = \frac{1}{V} \, \sum_{\bk,\alpha,\beta} |A_{\alpha \beta}(\bq,\bk)|^2 \; \frac{ 1-n_F\left(\xi_{\left(\frac{\bq}{2}+\bk \right) \, \alpha} \right) - n_F\left(\xi_{\left(\frac{\bq}{2}-\bk \right) \, \beta} \right)}{i q_l - \xi_{\left(\frac{\bq}{2}+\bk \right) \, \alpha} - \xi_{\left(\frac{\bq}{2}-\bk \right) \, \beta}},$
%\eeq
with $n_F(x) = 1 / (e^{x/T} + 1)$ denoting the Fermi function, $A_{\alpha \beta}$ is the amplitude of the singlet in the two particle-state with momenta $\bq/2\pm\bk$ and helicities $\alpha$ and $\beta$, and $\xi \equiv \epsilon -\mu$. The analysis also produces the pairing susceptibility
$\chi(q) = L(q) \left(\frac{L(q)}{M(q)} -1 \right)$,
whose divergence from the positive side up on the reduction of  temperature indicates a pairing instability. The first such divergence of $\chi(0,\bq)$ occurs at $\bq=\bzero$ (as we have verified), i.e., the system is most susceptible to  homogeneous pairing. $T_c$ is then obtained via (the Thouless criterion\cite{Thouless1960})
%\beq\label{eqn:Thouless}
$-\frac{1}{4 \pi \as} - \frac{1}{4V} \sum_{\bk,\alpha} \left(\frac{1-2 \, n_F\left(\xi_{\bk \, \alpha}\right)}{\xi_{\bk \, \alpha}} -\frac{2 \;}{|\bk|^2} \right) = 0.$
%\eeq

The equation  of state of the system is determined from \eqn{eqn:Sgauss} as
\bea 
\textstyle{\rho(T,\mu)}   & = &  \textstyle{\frac{1}{V} \sum_{\bk,\alpha} n_F\left(\xi_{\bk \, \alpha} \right) }\non \\
& &  \textstyle{ - \frac{1}{V} \sum_{\bq} \frac{1}{\pi} \int_{-\infty}^{\infty} \D{\omega} \; n_B(\omega) \; \frac{\dou \; \text{arg} \left(M(\omega^+,\bq)\right)}{\dou \mu},} \mylabel{eqn:EOS}
\eea
where arg($z$) is the argument of the c-number $z$, $n_B(x) = 1 / (e^{x/T} - 1)$ is the Bose function, and $M(\omega^+,\bq)$ is the analytic continuation of $M(iq_\ell\rightarrow z,\bq)$ evaluated just above the real axis in the $z$-plane.  $\mu$ at a given $T$ is then determined from the solution of the equation $\rho(T,\mu) = \rho_0$. $M(z,\bq)$ may have an isolated zero (below the scattering threshold $\omega_0(\bq)$) at $z=\omega_b(\bq)$ along the real axis; this signals the presence of a bosonic bound state of a fermion pair with center of mass momentum $\bq$. It is useful to rewrite the equation of state by explicitly identifying the contributions to $\rho$ 
\beq\mylabel{eqn:rhosplit}
\rho(T,\mu) = \rhoF(T,\mu) + \rhop(T,\mu) + \rhobc(T,\mu)
\eeq
where $\rhoF(T,\mu) = \frac{1}{V} \sum_{\bk,\alpha} n_F\left(\xi_{\bk \, \alpha} \right) $ is the fermion contribution, $\rhop(T,\mu) = - \frac{1}{V} \sum_{\bq} n_B(\omega_b(\bq)) \frac{\dou \omega_b(\bq)}{\dou \mu}$ is the contribution from the bosonic poles, and $\rhobc(T,\mu) = - \frac{1}{V} \sum_{\bq} \int_{\omega_0(\bq)}^{\infty} \D{\omega} \; n_B(\omega) \; \frac{\dou \; \text{arg} \left(M(\omega^+,\bq)\right)}{\dou \mu} $ is the contribution from the scattering continuum manifested as a branch cut of $M(z,\bq)$ along the real $z$-axis. 

\begin{figure}
\centerline{\includegraphics[width=\myfigwidth]{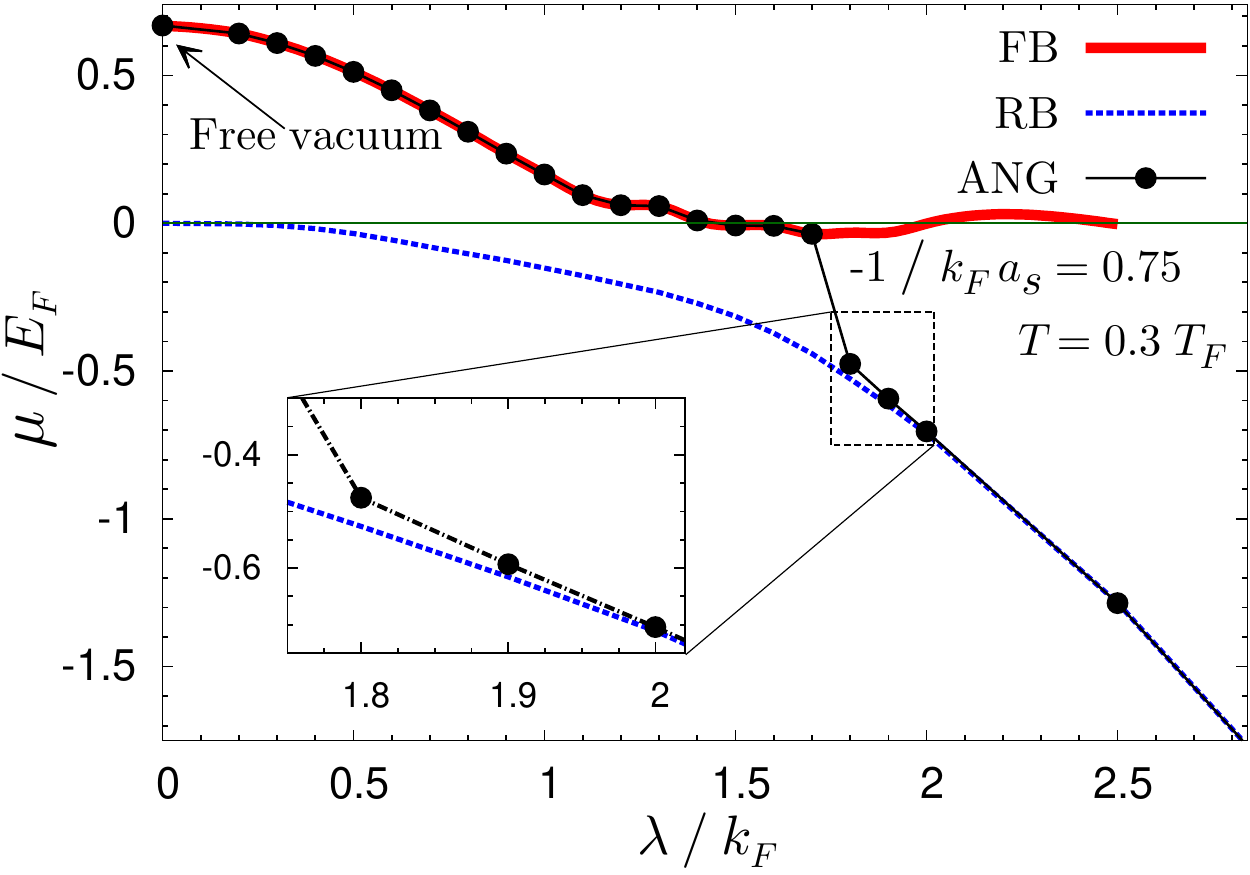}}
\caption{(Color online) {\bf Dependence of chemical potential ($\mu$) on RSOC strength:} The Gaussian theory has two distinct solutions for $\mu$,  called the free vacuum branch (FB) (higher $\mu$, solid line), and the rashbon branch (RB) (lower $\mu$, dashed line). RB always has the lower free energy within the Gaussian theory. The solid line with dots indicates $\mu$ in the approximate non-Gaussian (ANG) theory developed in this paper. Non-Gaussian effects eliminate RB for $\lambda \lesssim k_F$.
}
\mylabel{fig:mu_vs_lam}
\end{figure}

\noindent
{\bf Inadequacy of the Gaussian Theory:} The Gaussian theory,  notably successful\cite{Zwerger2012} in the description of the interacting Fermi gas in free vacuum, has rather peculiar features in the presence of RSOC ({\it non-Abelian} gauge field) of the type  $\blam=(\lambda_r,\lambda_r,\lambda_p)$ where $\lambda_r \ge \lambda_p$. While we focus on the spherical gauge field, our discussion will be applicable to all such gauge fields.

\Fig{fig:mu_vs_lam} shows the dependence of $\mu$ on $\lambda$ at a fixed low $T$ ($T < T_F$) and negative $\as$. The remarkable feature is that the Gaussian theory has {\it two} solutions for $\mu$ for a given set of parameters. For $\lambda \ll \kf$, one of the solutions, called the ``free vacuum branch''(FB) (see \Fig{fig:mu_vs_lam}), is smoothly connected to that of the free vacuum found in previous works.\cite{Drechsler1992,SaDeMelo1993}  The other solution always has $\mu < 0$, even for $\lambda \ll \kf$. For large $\lambda$, $\mu$ in this branch is determined by the rashbon dispersion, and hence called the rashbon branch (RB). Curiously $\mu$ along RB, which {\it always} has the lower free energy, approaches $0^-$ as $\lambda \rightarrow 0$.  This suggests that the equilibrium state of the Gaussian  theory with RSOC is not continuously connected to the free vacuum in the limit of $\lambda \rightarrow 0$! (see \Fig{fig:mu_vs_lam}.)

The physics of RB at small $\lambda$ can be traced to the contribution $\rhop$ from the bound bosonic states to the total density (\eqn{eqn:rhosplit}). \Fig{fig:wb_vs_q} shows the dispersion of such bosons as a function of their momentum $\bq$. Key points to be noted, whenever $\mu < 0$, are (i) even for negative scattering lengths, there are bound bosonic states in RB whenever $|\bq| < q_0(= \frac{2 \lambda}{\sqrt{3}})$, while they cease to exist at larger $|\bq|$. (ii) the binding energy of the bosons ($E_b(\bq) = \omega_0(\bq) - \omega_b(\bq)$), even though significant for small $|\bq|$, is vanishingly small in the range $\frac{q_0}{2} \lesssim |\bq| \le q_0$. This physics is quite similar to what is found in the two body problem.\cite{Vyasanakere2011Rashbon} Such a bosonic dispersion can therefore accommodate a large number of particles forcing $\mu$ to be self-consistently negative. Although $E_b(\bq=\bzero) \to 0$  for vanishingly small $\lambda$ this phenomenon persists, resulting in RB not being smoothly connected to the free vacuum. Note also that FB does not have any contribution from $\rhop$, since in this regime there is no bosonic bound state for $\mu >0$.

\begin{figure}
\centerline{\includegraphics[width=\myfigwidth]{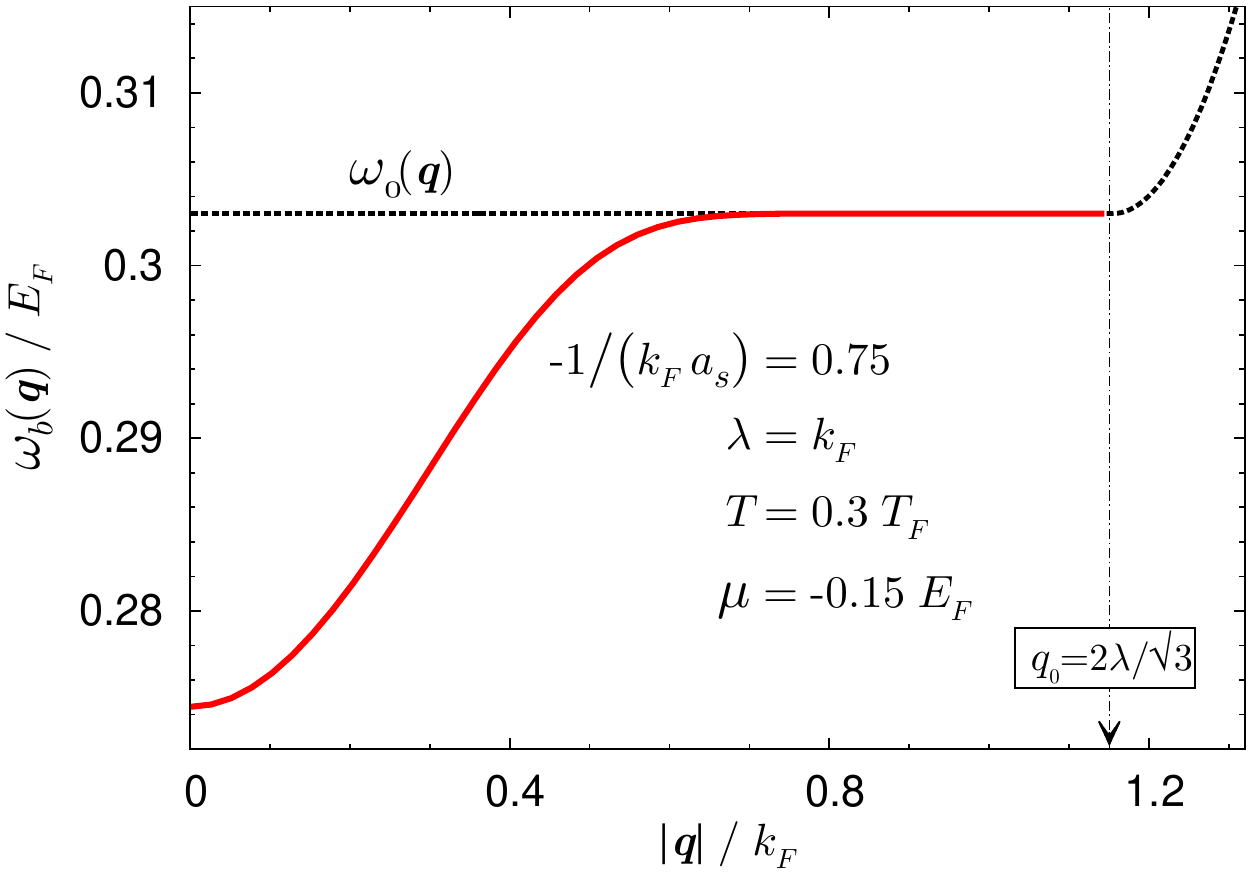}}
\caption{(Color online) {\bf Energy dispersion of the bound boson (fermion-pair) in the Gaussian theory:} The dashed black line is the scattering threshold $\omega_0(\bq)$ as a function of $|\bq|$ (momentum of the pair).}
% For a negative chemical potential in the weakly attractive regime ($\as < 0$), a well formed bosonic state exists when $|\bq| \ll q_0$, and no such state exists for $|\bq| \ge q_0$. In the regime $\frac{q_0}{2} \lesssim |\bq| \leq q_0$ the binding energy of the boson is vanishingly small.}
\mylabel{fig:wb_vs_q}
\end{figure}

As is evident, the fully formed bound states in the range $\frac{q_0}{2} \lesssim |\bq| \le q_0$ with a vanishing binding energy can easily be destabilized. In particular, the quartic term in \eqn{eqn:SExpansion} with the coupling $K$ describes the interactions between the pairing fluctuations $\eta(q)$. A natural question is whether these weakly bound states are stable when the interactions between the bosonic fluctuations are taken into account. We address this question by quantifying the strength of these beyond-Gaussian effects by the parameter $b$ (proportional to $K(q_1,q_2;q_3,q_4)$ in the limit of zero momentum\cite{Drechsler1992,Randeria1995}) obtained as
\beq\mylabel{eqn:bDef}
\textstyle{b =  \frac{1}{4V} \sum_{\bk,\alpha} \left(\frac{1-2 \, n_F\left(\xi_{\bk \, \alpha}\right)}{{\xi_{\bk \, \alpha}}^3} - \frac{2 \, n_F\left(\xi_{\bk \, \alpha}\right)\left(1-n_F\left(\xi_{\bk \, \alpha}\right)\right)}{T \, {\xi_{\bk \, \alpha}}^2} \right)}.
\eeq
When $\mu$ is large and negative, as in a ``boson dominated'' state where the most prominent contribution arises from $\rho_b$, $b\approx\frac{\lambda^2 - 2\mu}{32 \pi \sqrt{2} (-\mu)^{5/2}}$. The physical meaning of $b$ can be made evident by noting that $b \sim {a_{BB}}^3$ when $\lambda =0$ and the scattering length is small positive (free vacuum BEC side). Here $a_{BB}$ is the scattering length of two bosons (bound fermion pairs) and is proportional to $\as$.\cite{Hu2006} Furthermore, for any $\as$, as $\lambda \to \infty$, $b \to \lambda^{-3}$, which can be immediately identified with ${a_{RR}}^3$, where $a_{RR}$ is the rashbon-rashbon scattering length.\cite{Vyasanakere2012Collective} Therefore, $b^{1/3}$, is a length scale that characterizes the interactions among the pairing fields. Interestingly, this parameter is nonzero in the limit of $\lambda \to 0^+$ and {\it grows} with increasing $\lambda$ attaining a peak when $\lambda \approx k_F$(see inset of \Fig{fig:Fig_3_lam_c}), subsequently possessing the just discussed asymptotic behavior at large $\lambda$.

\begin{figure}
\centerline{\includegraphics[width=\myfigwidth]{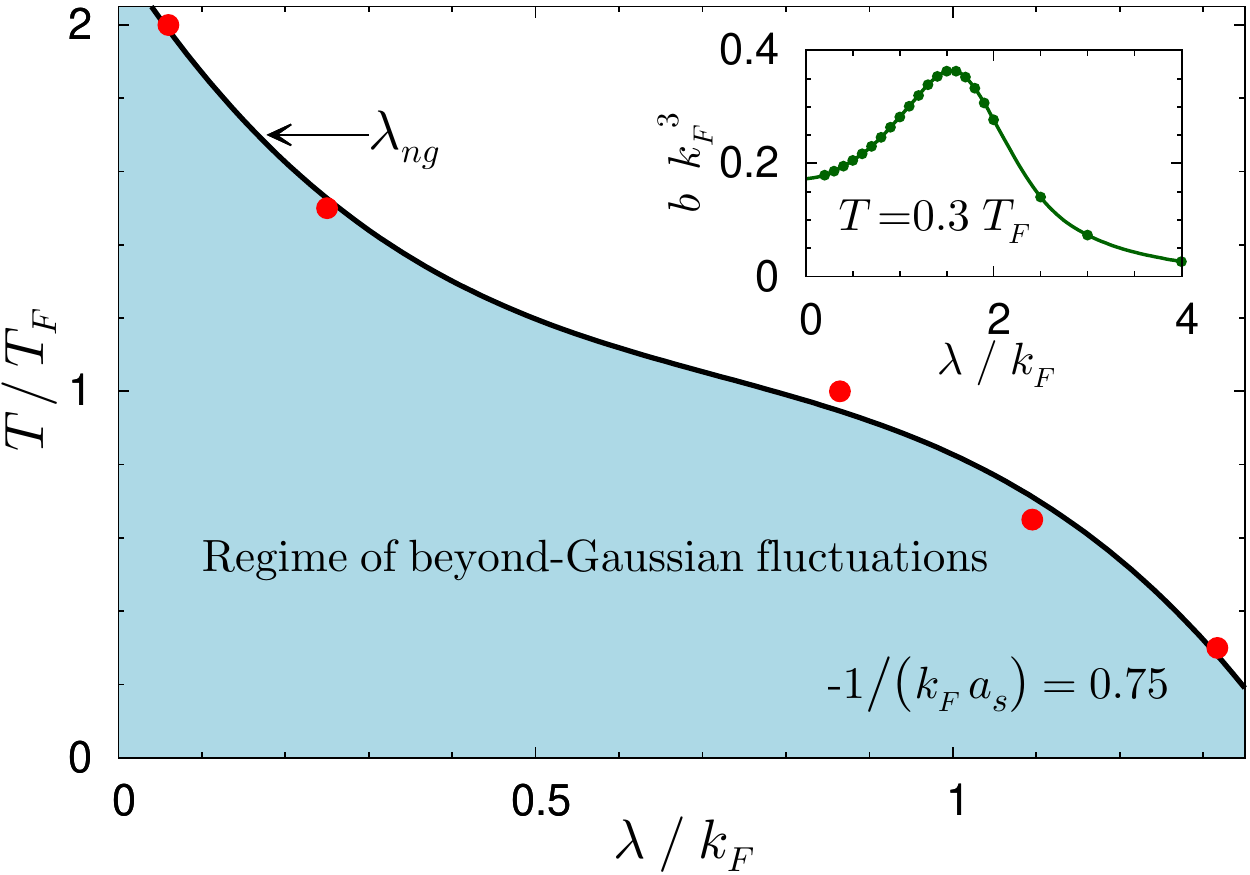}}
\caption{(Color online) {\bf Regime where non-Gaussian effects are ineluctable:} At a given temperature $T$ and scattering length $\as$, non-Gaussian effects play a crucial role when $\lambda \lesssim \lambda_{ng}(T,\as)$. Inset shows the parameter $b$ (see text) that characterizes the non-Gaussian effects at a fixed temperature.}
\mylabel{fig:Fig_3_lam_c}
\end{figure}

The effects of $b$ on  the weakly bound states can now be estimated in a physical manner. The lowest order effect of $b$ would be to shift the energy of the bound state via a Hartree shift, i.e., $\omega_b(\bq) \to \omega_b(\bq) + \kappa b^{1/3} \rho_b(T,\mu)$ where $\kappa$ is a dimensionless number of order unity\footnote{\FNkappa}. Clearly, the bound bosonic state will be unstable if the shift takes it into the scattering continuum, i.e., a necessary condition for the stability of the bound state is that $\omega_b(\bq) + \kappa b^{1/3} \rho_b(T,\mu) \le \omega_0(\bq)$. Thus RB can be stable only if
\beq\mylabel{eqn:Stability}
\textstyle{E_b(\bq=\bzero) \ge  \kappa b^{1/3} \rho_b(T,\mu).}
\eeq
Using this criterion we  obtain the regime (see \Fig{fig:Fig_3_lam_c}) where RB is eliminated by non-Gaussian effects i.e., $\lambda \le \lambda_{ng}(T,\as)$. For a given $\as$, we find that $\lambda_{ng}$ increases with decreasing temperature\footnote{\FNlamNGbcs}. These estimates provide a lower bound of $\lambda_{ng}$, which results in $\lambda_{ng} \approx k_F$ for temperatures $T \lesssim T_F$. Thus beyond-Gaussian effects are {\it crucial} in the most interesting regime of parameters. 

\noindent
{\bf Approximate Non-Gaussian Theory:} Having firmly established that even a qualitatively correct description of spin-orbit coupled Fermi gases necessarily requires a beyond-Gaussian theory, we propose and discuss one such theory. A key desideratum of such a theory is the elimination of RB for $\lambda \lesssim \lambda_{ng}$, and a smooth evolution (at  given $T,\as$) from the free vacuum state at  vanishing $\lambda$ to the rashbon gas at large $\lambda$. The implementation of such a theory is a formidable challenge, even as we note that Gaussian theory itself requires considerable calculational effort\footnote{\FNCalcEffort}. Faced with this reality, we develop an approximate non-Gaussian (ANG) theory by a suitable modification to the equation of state (\eqn{eqn:EOS}) that only entails the same calculational complexity as the Gaussian theory.
The approximation follows the physical argument that the non-Gaussian term  $b$ shifts $\omega_b(\bq)$ to $\omega_b(\bq) + \kappa b^{1/3} \rho_b^{GS}$ where $\rho_b^{GS}$ is the bound boson contribution calculated within the Gaussian approximation. Only those bosonic states that remain below the scattering continuum after this energy shift, i.e., the bosonic states for all $|\bq| \le q_b$ obtained by $E_b(q_b) = \kappa b^{1/3} \rho_b^{GS}$, are stable to non-Gaussian effects. These arguments provide for the approximation to the contribution of the bosonic bound pairs to the equation of state (\eqn{eqn:EOS}) as $\rho^{\text{ANG}}_b(T,\mu) =  - \frac{1}{V} \sum_{|\bq| \le q_b} n_B(\omega_b(\bq)) \frac{\dou \omega_b(\bq)}{\dou \mu}$. 

\Fig{fig:mu_vs_lam} shows the results of this approximate non-Gaussian theory (see the curve marked ANG). The approximate theory does indeed possess the key features desired, notably the elimination of RB for $\lambda \ll k_F$. In this regime, the ANG chemical potential smoothly connects to free vacuum value. Furthermore, when $\lambda \gg k_F$, the ANG recovers the rashbon gas. 
In the ANG theory, RB appears only after a particular value of $\lambda$ which depends on $T$ and $\as$ at which the solution switches from FB to RB. 
This evolution should be smooth in a detailed  theory which also includes non-Gaussian effects in the free vacuum.

\begin{figure}
\centerline{\includegraphics[width=\columnwidth]{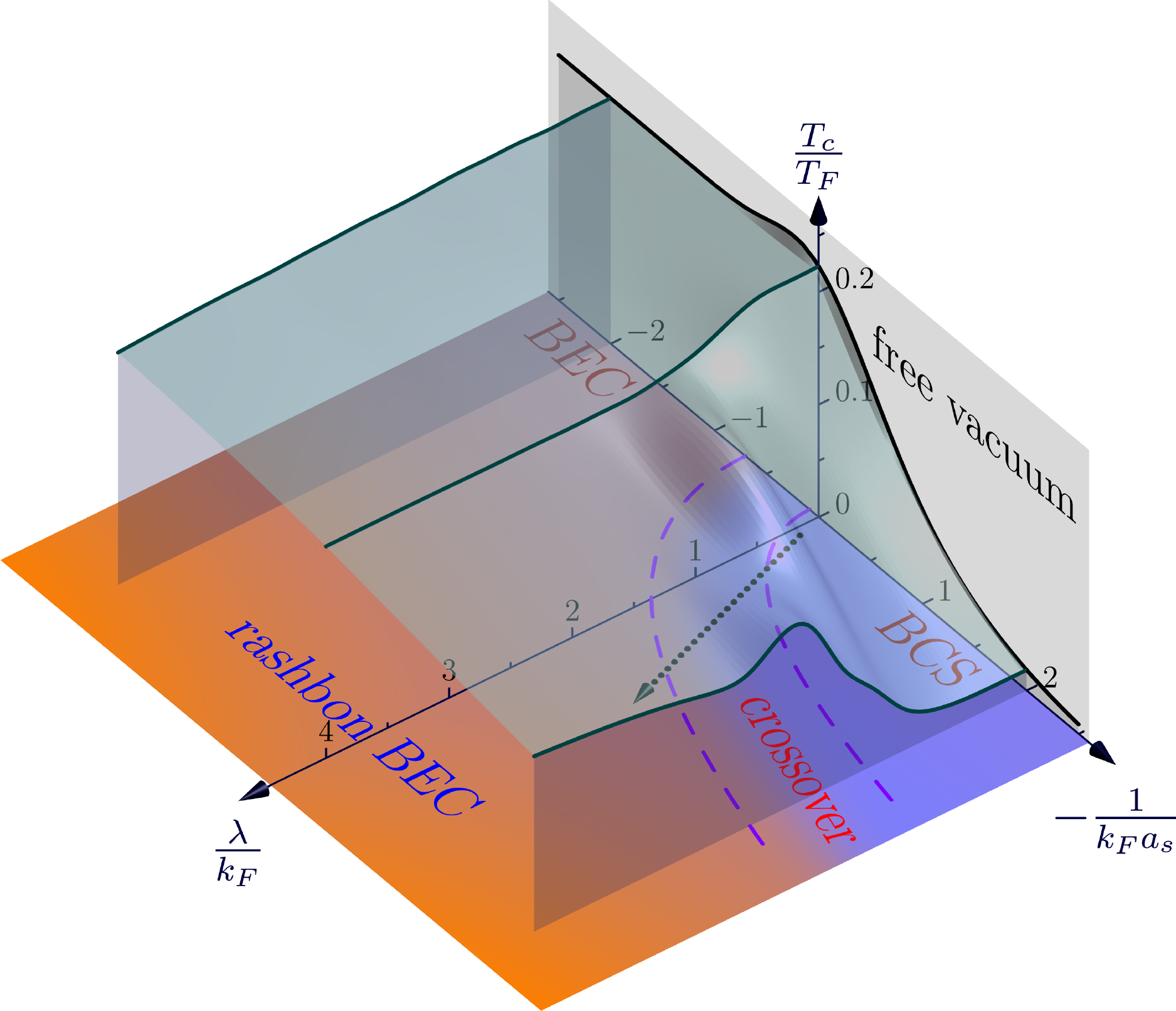}}
\caption{(Color online) {\bf Phase diagram:} Dependence of superfluid $T_c$ on $\as$ and $\lambda$ (ANG theory). 
%For a small positive scattering length, the $T_c$ evolves gradually from that of the usual BEC of fermions to that of the rashbon-BEC. On the other hand, for weak attractions (small negative scattering lengths) the $T_c$ is enhanced by the spin orbit coupling. For large $\lambda$, the $T_c$ evolves to that of the rashbon-BEC irrespective of the scattering length. 
Crossover occurs in regime enclosed by the dashed lines. The dotted line indicates a candidate path that can be traced out in a cold atoms experiment by varying the density at a fixed scattering length and RSOC. 
}
\mylabel{fig:Fig_PD}
\end{figure}

\noindent
{\bf Phase Diagram:} We now use the ANG theory to obtain the  phase diagram shown in \Fig{fig:Fig_PD}. For a small positive
scattering length, which obtains a BEC of fermion pairs in 
free vacuum, increasing RSOC engenders a smooth
crossover to the rashbon-BEC with the $T_c$ gradually changing
from that set by the free vacuum boson mass (twice the fermion
mass) to that set by the rashbon mass. At the resonant
scattering length, the $T_c$ again evolves from that of the
free vacuum unitary Fermi gas, to that of the rashbon-BEC.
The scenario for a small negative scattering length is significantly different as discussed below.

One of the key aspects of the phase diagram \Fig{fig:Fig_PD}, shown in detail in
\Fig{fig:Fig_NM}, is the large enhancement of $T_c$ for a system with a weak attractive
interaction. For example, for ${-1\over\kf\as}=2$, the $T_c$ is
enhanced from $0.02T_F$  at $\lambda = 0$ to about $0.1 T_F$ when $\lambda \approx \kf$. Further, there is a regime of $\lambda$ where 
 $T_c$ decreases.  Beyond this, $T_c$ is determined by two-body physics
 as shown by dashed-dot lines in
\Fig{fig:Fig_NM}. \Fig{fig:Fig_NM} also
shows the mean field $T_c$ which makes  fluctuation
effects evident. For example, the $T_c$ from the ANG theory is
about $85\%$ of the mean field $T_c$ for $-\frac{1}{\kf\as} =
2.0$, and it is reduced to about $60\%$ of the mean field
value when $-\frac{1}{\kf\as} = 0.75$. The enhancement of $T_c$ for weak attractive interactions indicated by our ANG theory is a remarkable feature of spin orbit coupled systems ({\it non-Abelian} gauge fields).  Finally, for any given $\as$ (including $\as < 0$), the $T_c$ at large $\lambda$ ($\lambda \gtrsim (\kf,1/|\as|)$) is independent of $\as$, determined only by the rashbon mass.

\begin{figure}
\centerline{\includegraphics[width=\myfigwidth]{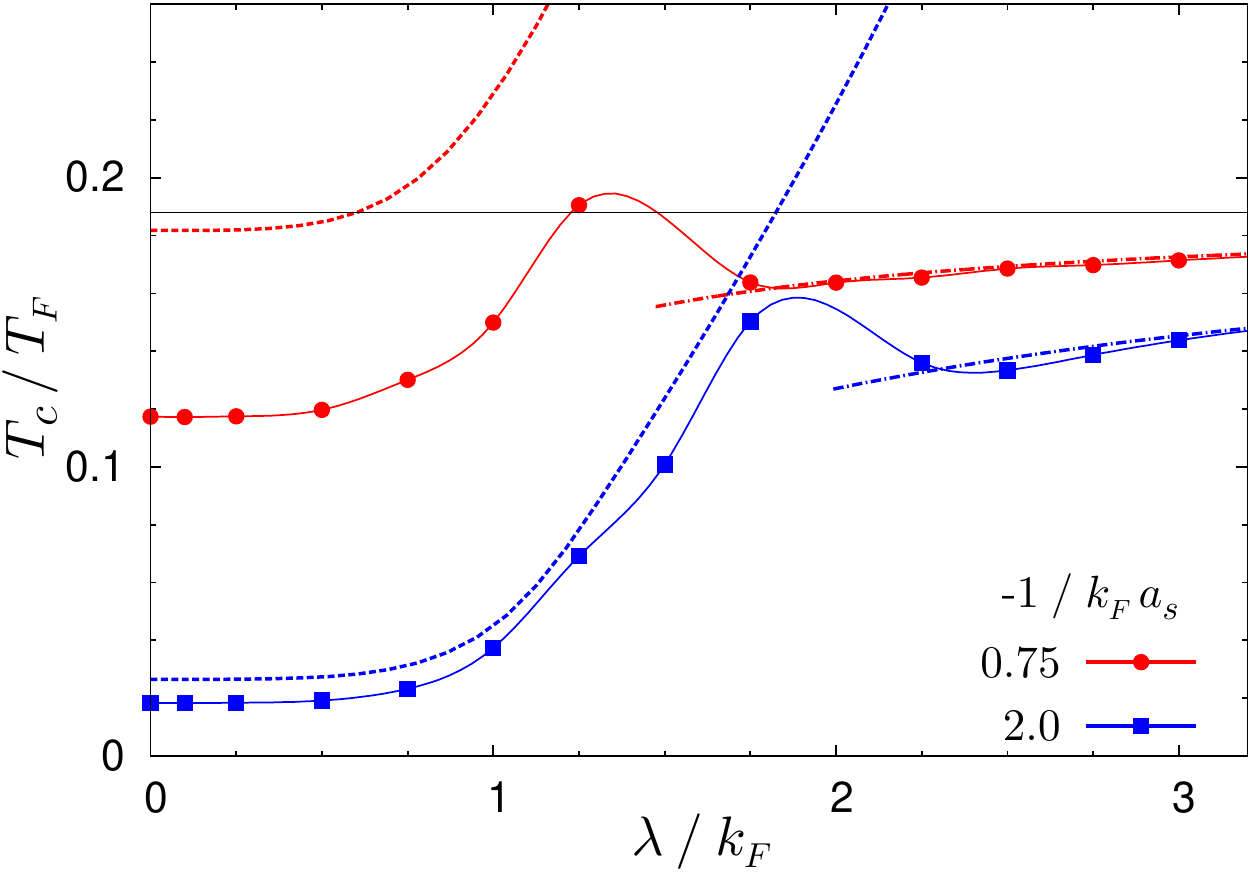}}
%\centerline{{\bf (a)~~~~~~~~~~~~~~~~~~~~~~~~~~~~~~~~~~~~~~~~~~~~~~~~~~~~~~~~~~~~~~~~~~~~~~~~(b)}}
\caption{(Color online) {\bf Non-monotonic dependence of $T_c$ on $\lambda$ for weak attraction:} Points indicate the $T_c$ calculated from our  ANG theory (lines through the points are a guide to the eye). Dashed lines -- $T_c$ from mean field theory. Dashed-dot lines -- $T_c$ estimated from the condensation of the bound-fermion pairs. Thin horizontal line -- $T_c$ of the rashbon-BEC.
}
\mylabel{fig:Fig_NM}
\end{figure} 

As shown in \Fig{fig:Fig_PD}, there is a much bigger regime of parameters (with weak interactions and RSOC) over which the crossover from a BCS like ground state to a rashbon-BEC occurs. The central point is that the superfluid with high $T_c$ occurs in this crossover regime. Indeed, it will be interesting to mimic this crucial finding in material systems to provide routes to making superconductors with high transition temperatures. On a different token, this physics can be uncovered in a cold atoms experiment at fixed negative $\as$ and RSOC, by working with different trap centre densities, tracing out a path akin to the dotted line shown in \Fig{fig:Fig_PD}. Another interesting point to note is that the enhanced binding induced by the RSOC will result in significant pseudogap features\cite{Gaebler2010} which could be observed even at higher temperatures.

In summary, we have shown the crucial role of beyond-Gaussian effects in spin orbit coupled Fermi gases. We have developed a simple theory that incorporates the beyond-Gaussian effects in an approximate fashion. Using this theory we obtain the phase diagram of the system. A key result of our calculation is the demonstration of the enhancement of the exponentially small superfluid transition temperature with weak attraction to values comparable to Fermi temperature. This important point provides clues to producing superconductors with high transition temperatures. Our approximate non-Gaussian theory uncovers the rich physics in spin-orbit coupled gases providing motivation for further detailed theoretical considerations. Promising routes to treat beyond-Gaussian effects include the $G_0-G$ or $G-G$ schemes.\cite{Chen2005,Strinati2012}

\noindent
{\bf Acknowledgements:} JV acknowledges support from CSIR, India.  VBS is
grateful to DST, India (Ramanujan grant), DAE, India (SRC grant) and IUSSTF
for generous support.

\bibliography{refNSR_nagf}

%UCOMMENT THIS TO GET PRL FORMAT
%\end{document}

\clearpage

\setcounter{page}{1}
\newpage

\appendix

\addto{\captionsenglish}{\renewcommand*{\appendixname}{}}

\centerline{{\bf Supplemental Material}}
\centerline{for}
\centerline{{\bf \OurTitle}}
\centerline{by Jayantha P. Vyasanakere and Vijay B. Shenoy}
\bigskip

\noindent
\section{Fluctuation Theory Formulation}

Many body physics is studied by introducing  four component Nambu fields $\varPsi^\star(k) = \left( c^\star_+(k) \,\, c_+(-k) \,\, c^\star_-(k) \,\, c_-(-k) \right)$, where $k$ denotes the four-vector $(ik_n, \bk)$, $ik_n$ being a Fermi-Matsubara frequency. The action of a system of interacting fermions is made up of three pieces:
\beq\mylabel{eqn:TotalAction}
{\cal S}[\varPsi] = {\cal S}_0[\varPsi]  + {\cal S}_\upsilon[\varPsi]  + {\cal S}_F[\varPsi] .
\eeq
The term ${\cal S}_0$ describes the kinetic energy including the RSOC induced by the non-Abelian gauge field,
\beq\mylabel{eqn:S0}
{\cal S}_0[\varPsi]  = \half \sum_k\varPsi^\star(k) (-G_0^{-1}(k,k')) \varPsi(k').
\eeq
where $G_0^{-1}(k,k') = \mbox{Diag}(ik_n - \xi_{\bk \, +},ik_n + \xi_{\bk \, +},ik_n - \xi_{\bk \, -},ik_n + \xi_{\bk \, -}) \delta_{k,k'}$ with $\xi_{\bk \, \alpha} = \varepsilon_{\bk \, \alpha} - \mu$ and $\mu$ is the chemical potential.
The second term, ${\cal S}_\upsilon$, in the action (\eqn{eqn:TotalAction}) describes the contact attraction among fermions as
\beq\mylabel{eqn:Supsilon}
{\cal S}_\upsilon[\varPsi]  = \frac{\upsilon T}{V} \sum_q S^\star(q) \, S(q) ,
\eeq
where  $T$ and $V$ are respectively the temperature and  volume of the system, $\upsilon$ is the bare interaction parameter. This last quantity  is traded for the $s-$wave scattering length $\as$ through regularization as $ \frac{1}{4 \pi \as} = \frac{1}{\upsilon} + \Lambda $, where $\Lambda = \frac{1}{V} \sum_{\bk}\frac{1}{k^2}$  denoting the ultraviolet cutoff. The quantity
$S^\star(q) = \sum_{k,\alpha \beta} A_{\alpha \beta}(\bq,\bk) \, c^\star_{\alpha}\left(\frac{q}{2} + k\right) \, c^\star_{\beta}\left(\frac{q}{2} - k\right)$ stands for the  singlet pair density in Matsubara-Fourier space, $q =(iq_\ell, \bq)$, where $iq_\ell$ is the Bose-Matsubara frequency, $\bq$ is the center of mass momentum and $\bk$ is the relative momentum of a two-particle state with particles having  helicities $\alpha$ and $\beta$. $A_{\alpha \beta}(\bq,\bk)$ is the weight of such a state in the singlet sector.  The third term in  \eqn{eqn:TotalAction} contains external pairing source fields $F(q)$, 
\beq\mylabel{eqn:SF}
{\cal S}_F[\varPsi]  = \sqrt{\frac{T}{V}} \, \sum_q F(q) S^\star(q) + F^\star(q) S(q).
\eeq
This term anticipates a pairing instability in the system, and is added solely to aid the calculation of the pairing susceptibility (most of the formulae, therefore, will have $F=0$).

We now perform a Hubbard-Stratanovich transformation on ${\cal S}_\upsilon$  by introducing pairing fields $\eta(q)$,
\beq\mylabel{eqn:HS}
{\cal S}[\varPsi,\eta,F] = \sum_{k,k'} \varPsi^\star(k) (-G^{-1}(k,k')) \varPsi(k') -\frac{1}{\upsilon} \sum_q \eta^\star(q) \, \eta(q) \, , 
\eeq
where
\beq
G^{-1}(k,k') = G_0^{-1}(k,k') - \bgam(k,k'),
\eeq 
\beq
\bgam(k,k') = \left(
\begin{array}{cccc}
 0 & \gamma_{++}(k,k') & 0 & \gamma_{+-}(k,k') \\
 \tilde{\gamma}_{++}(k,k') & 0 &  \tilde{\gamma}_{+-}(k,k') & 0 \\
0 & \gamma_{-+}(k,k') &  0 & \gamma_{--}(k,k') \\
\tilde{\gamma}_{-+}(k,k') & 0 & \tilde{\gamma}_{--}(k,k') & 0  
\end{array}
 \right)
\eeq
with
\begin{small}$\gamma_{\alpha\beta}(k,k') = \sqrt{\frac{T}{V}} \, \sum_q \gamma(q) A_{\alpha \beta}\left(\bq,\bk - \frac{\bq}{2}\right) \delta_{q, k-k'}$, $
\tilde{\gamma}_{\alpha \beta}(k,k') = \sqrt{\frac{T}{V}} \, \sum_q \gamma^*(-q) A^{*}_{\beta \alpha}\left(-\bq,\bk - \frac{\bq}{2}\right) \delta_{q, k-k'}$\end{small}, and 
$\gamma(q) = \eta(q) + F(q)$.
The action is now quadratic in fermionic fields which can be integrated to yield
\beq\mylabel{eqn:SetaFSM}
{\cal S}[\eta,F] = - \ln\det[-G^{-1}] - \frac{1}{\upsilon} \sum_q \eta^\star(q) \, \eta(q).
\eeq
This is \eqn{eqn:SetaF} in the main text.

\end{document}